\documentclass[amssymb,showpacs,amsmath,nobibnotes,nofootinbib,aps,prb,superscriptaddress,twocolumn]{revtex4}
\usepackage{graphicx,dcolumn,bm}
\usepackage[dvips]{color}
\begin{document}
\title{Kagom\'{e} staircase compound $\rm Co_3V_2O_8$ in an applied magnetic field: single-crystal neutron-diffraction study} 
\newcommand{\NVO}{$\rm Ni_3V_2O_8$}
\newcommand{\CVO}{$\rm Co_3V_2O_8$}
\newcommand{\MVO}{$\rm M_3V_2O_8$}
\newcommand{\Kag}{kagom\'{e}}
\newcommand\Tn{$T_{N}$}
\newcommand\afm{antiferromagnet}
\newcommand\PRB[3]{Phys. Rev. B {\bf {#1}}, {#2} ({#3})}
\newcommand\PPRL[3]{Phys. Rev. Lett. {\bf {#1}}, {#2} ({#3})}
\newcommand\JJPCM[3]{J. Phys.: Condens. Matter {\bf {#1}}, {#2} ({#3})}
\newcommand\JJPSJ[3]{J. Phys. Soc. Japan {\bf {#1}}, {#2} ({#3})}
\newcommand\JJMMM[3]{J. Mag. and Mag. Mater. {\bf {#1}}, {#2} ({#3})}
\newcommand\PhysB[3]{Physica B {\bf {#1}}, {#2} ({#3})}  
\author{O.A.~Petrenko}
\author{N.R.~Wilson}
\author{G.~ Balakrishnan}
\author{D.~$\rm M^cK$~Paul}
\affiliation{Department of Physics, University of Warwick, Coventry, CV4 7AL, UK}
\author{G.J.~McIntyre}
\affiliation{Institut Laue-Langevin, 6 rue Jules Horowitz, BP 156 - 38042 Grenoble Cedex 9, France}
\date{\today}

\begin{abstract}
The magnetic properties of \CVO\ have been studied by single-crystal neutron-diffraction.
In zero magnetic field, the observed broadening of the magnetic Bragg peaks suggests the presence of disorder both in the low-temperature ferromagnetic and in the higher-temperature antiferromagnetic state.
The field dependence of the intensity and position of the magnetic reflections in \CVO\ reveals a complex sequence of phase transitions in this \Kag\ staircase compound.
For $H\parallel a$, a commensurate-incommensurate-commensurate transition is found in a field of 0.072~T in the \afm ic phase at 7.5~K.
For $H\parallel c$ at low-temperature, an applied field induces an unusual transformation from a ferromagnetic to an \afm ic state at about 1~T accompanied by a sharp increase in magnetisation.
\end{abstract}
\maketitle

The family of transition metal vanadates \MVO\ with M being Co, Ni, Cu, and Mn has been intensively studied in the context of frustrated magnetism,\cite{Rogado_SSC_2002,Rogado_JPCM_2003,Lawes_PRL_2005,Chen_PRB_2006,Morosan_PRB_2007} since the magnetically coupled ions in these compounds form a network reminiscent of the \Kag\ lattice.
The chemical bonding between the magnetic M$^{2+}$ ions (the V$^{5+}$ ions remain nonmagnetic) forces the magnetic layers to buckle into a staircase formation, the so-called \Kag\ staircase lattice.
Although the magnetic properties of different members of the family vary significantly, a unifying feature for all the members is an extreme richness of their highly anisotropic magnetic phase diagrams.\cite{Morosan_PRB_2007,Wilson_JPCM_2007,Lawes_PRL_2004}
The abundance of magnetic phases is attributed to a close proximity in energy of several competing states, with the selection of the ground state being readily influenced by a subtle balance amongst the nearest and further-neighbour exchange interactions, magnetic anisotropy, and an applied magnetic field.
An additional interest in \NVO\ is associated with the discovery of magnetically driven ferroelectric order in this compound,\cite{Lawes_PRL_2005} while studies of the magnetic ordering in $\beta$-$\rm Cu_3V_2O_8$ with $S=1/2$ are justified by the possible influence of quantum effects.\cite{Rogado_JPCM_2003}

In \CVO\ (CVO) a magnetic ordering at 11.3~K is seen as a relatively small peak in the heat capacity and a small kink in the magnetic susceptibility,\cite{Rogado_SSC_2002} while a further magnetic transition at 6~K is accompanied by a much more pronounced peak in the $C(T)$ curve and a significant rise in $\chi(T)$.\cite{Rogado_SSC_2002}
By using neutron-diffraction measurements, Chen {\it et al.}~\cite{Chen_PRB_2006} have shown that the transition at 11.3~K is from a paramagnetic to an \afm ic (AFM) incommensurate state and that the transition at 6~K is to a ferromagnetic (FM) state.
A number of lock-in transitions associated with incommensurate-to-commensurate phases have also been observed at intermediate temperatures.\cite{Chen_PRB_2006}
A peculiar feature of the FM state is that the size of the moment on one of the two Co$^{2+}$ sites, the so-called cross-tie site, is considerably reduced compared to the fully polarized state.\cite{Chen_PRB_2006,Wilson_PRB_2007}
The application of a magnetic field in this phase is found to enhance rapidly the cross-tie site magnetic moment, which reaches the expected value of 3~$\mu_B$ in higher fields.\cite{Wilson_PRB_2007}
\begin{figure}[tb] 
\begin{center}
\includegraphics[width=0.6\columnwidth]{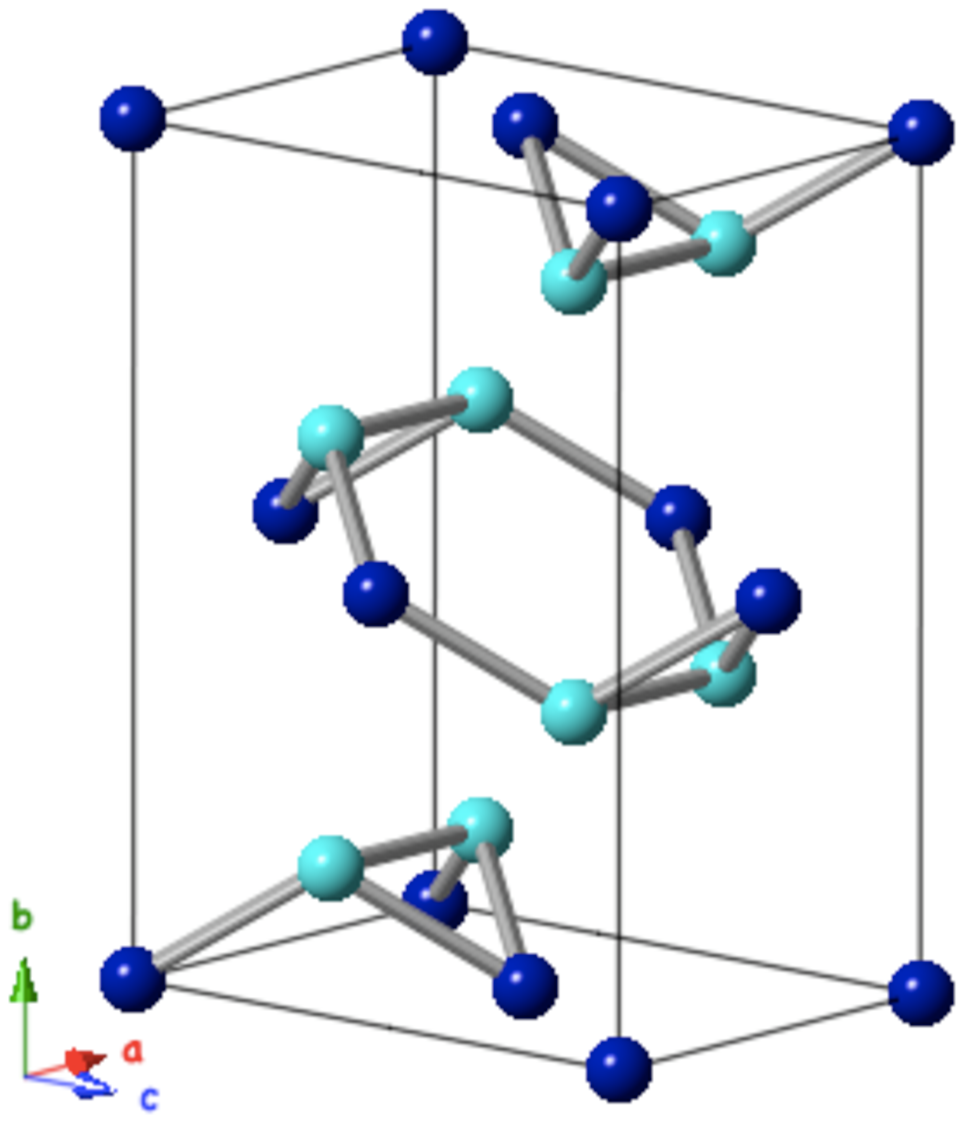}
\caption[]{(Colour online) The positions of the magnetic Co$^{2+}$ ions in the orthorhombic unit cell of \CVO\ (space group $Cmca$).
                  The cross-tie sites are shown in dark blue, and the spine sites are in light blue.}
\label{fig_structure}
\end{center}
\end{figure}

In an applied magnetic field, CVO demonstrates highly anisotropic behaviour both in terms of the absolute values of susceptibility~\cite{Balakrishnan_JPCM_2004} and in terms of the sequence of the field-induced phase transitions.\cite{Wilson_JPCM_2007,Yasui_JPSJ_2007}
Although the previous reports based on different experimental methods agree in general, they disagree over the detailed description of the behaviour of CVO in an applied magnetic field and in particular on the interpretation of the different magnetic phases.\cite{Wilson_JPCM_2007,Yasui_JPSJ_2007,Szymczak_PRB_2006,Kobayashi_JMMM,Yen_PhysB_2008}

Despite the recent progress in probing the magnetic interactions by inelastic neutron scattering~\cite{Ramazanoglu_PRB_2009} a unified view of the strength and signs of all the relevant magnetic forces has yet to be reached, as none of the proposed models is capable of explaining the sequence of phase transitions observed in CVO.
This absence of a generally accepted model is not particularly surprising, as a full description should incorporate 12 Co$^{2+}$ ions on two different sites in the unit cell (see Fig.~\ref{fig_structure}) in the presence of significant orbital effects (the rather small variation of energy with $q$ of the observed spin-waves compared to the gap suggests that the strength of exchange interactions in CVO is comparable with the single-ion anisotropy~\cite{Wilson_JMMM_2007}).
The matter is complicated further by the possibility of nonzero field-induced moments on the vanadium and oxygen ions.\cite{Qureshi_PRB_2009}

We have performed single-crystal neutron-diffraction measurements for two different directions of an applied magnetic field.
For $H\parallel a$, an unusual transition is found at intermediate temperatures.  
For $H\parallel c$, the observation of a field-induced AFM state implies a significant influence of the inter-\Kag -plane interactions on the magnetic structure formation.
The measurements have also shown that not all of the explored magnetic phases of CVO are long-range in their nature and that a significant degree of magnetic disorder is often present.

A CVO single-crystal sample of size 4$\times$2$\times$2~mm$^3$ was prepared as described previously.\cite{Balakrishnan_JPCM_2004}
Single-crystal diffraction was performed on the D10 diffractometer at the Institut Laue-Langevin, Grenoble, France.
An $80 \times 80$~mm$^2$ two-dimensional microstrip detector was used in the diffraction configuration.
Measurements were performed with $\lambda=1.53$~\AA\ and $\lambda=2.36$~\AA.
A pyrolitic-graphite monochromator and filter were used for the longer wavelength, while for the shorter wavelength a Cu (200) monochromator was used, still with the graphite filter to reduce the small half-wavelength contamination.
A vertical magnetic field of up to 2.5~T supplied by a standard cryomagnet was applied either along the $a$ axis or along the $c$ axis limiting the observable scattering to the planes  $(0kl)$ and $(hk0)$ respectively.
The measurements were performed by either ramping the magnetic field at constant temperatures and summing up the counts in a small area of the detector surrounding the reflection to obtain its peak intensity or by making $\omega$~scans at various fixed values of field/temperature.

\begin{figure}[tb] 
\begin{center}
\includegraphics[width=0.83\columnwidth]{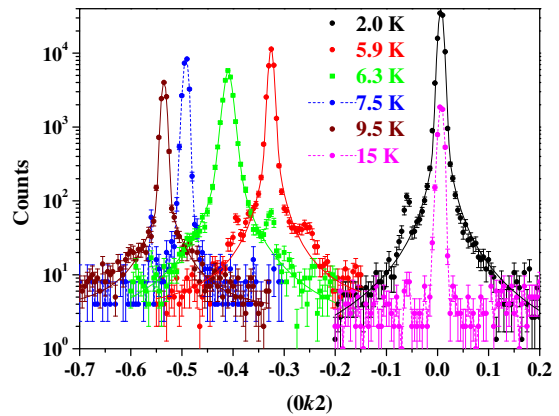}
\caption[]{(Colour online) Temperature dependence of neutron-diffraction profiles of scans along the $(0k2)$ direction in zero field.
		The measurements were performed with the use of a vertically focussing PG analyser.
		Solid lines are the fits consisting of two components, a main narrow Gaussian component and a less intense, broad Lorentzian component.
		Apart from the peak at $T=6.3$~K the Gaussian component is resolution limited.
		The data at 7.5~K and 15~K are not fitted, as they have only the Gaussian resolution limited component.}
\label{fig_analyser_Hzero}
\end{center}
\end{figure}
                  
The temperature evolution of the magnetic ordering in CVO was initially studied by collecting the intensity of the $(0k2)$ scans at different temperatures in zero field.
In general agreement with previously reported single-crystal~\cite{Chen_PRB_2006,Yasui_JPSJ_2007} and powder~\cite{Wilson_PRB_2007} neutron-diffraction data, the sequence of the magnetic transitions on cooling in zero field is from a paramagnetic phase to a higher-temperature AFM incommensurate phase (around 11.3~K), to a commensurate AFM $k=\frac{1}{2}$ phase (just below 9~K), to a lower-temperature AFM incommensurate phase (just above 7.5~K), to a commensurate AFM $k=\frac{1}{3}$ phase (stable in a narrow region between 5.8 and 6.1~K) and, finally, to a FM phase, which is stable below 5.8~K.

More accurate measurements of the positions and shapes of the magnetic reflection performed at different temperatures with an analyser in place of the position sensitive detector are shown in Fig.~\ref{fig_analyser_Hzero}.
In the paramagnetic region, at $T=15$~K, the $(002)$ peak has a nearly perfect Gaussian shape (apart from a small satellite, less then 0.3\% of the intensity of the main peak), which is limited by the instrumental resolution.
The same peak in the FM region has pronounced non-Gaussian tails, which amount to approximately 10\% of the integrated intensity (see Fig.~\ref{fig_analyser_Hzero}) and imply the presence of magnetic disorder throughout the sample.
The  $k=\frac{1}{3}$ AFM peak at  5.9~K and the incommensurate peak at 9.5~K also exhibit a dominant resolution-limited component and a much broader component, which could be adequately approximated by a Lorentzian shape, although very small additional reflections on top of the Lorentzian cannot be completely ruled out.
The main component of the incommensurate AFM peak at  $T=6.3$~K is not resolution limited, its width is estimated as 0.0151(5)~r.l.u. along the $(0k2)$ direction, which indicates that the magnetic correlation length along the $b$ axis does not exceed 220~\AA\ at this temperature. 
\begin{figure}[tb]
\begin{center}
\includegraphics[width=0.83\columnwidth]{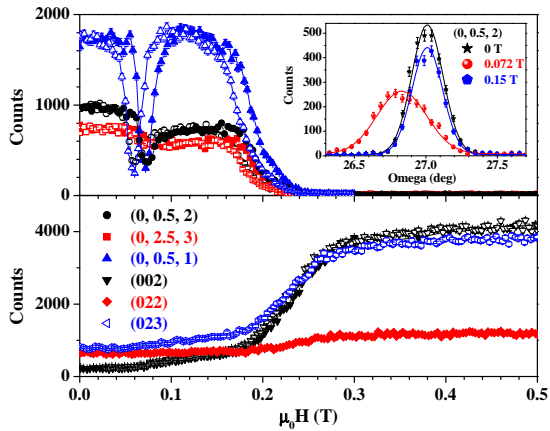}
\caption[]{(Colour online) Magnetic field dependence of the integrated intensity of several AFM (top panel) and FM (bottom panel) peaks in \CVO\ at $T=7.5$~K  for $H\parallel a$.
                  Solid/open symbols denote increasing/decreasing magnetic field ramps.
                  The inset shows $\omega$~scans for the $(0\frac{1}{2}2)$ reflection at 7.5~K in zero field, 0.072~T and 0.15~T.
                  Solid lines are Gaussian fits.
                  The FWHMs of the peaks are 0.26$^\circ$, 0.41$^\circ$, 0.25$^\circ$ respectively.}
\label{fig_field_scans_7p5K_Ha}
\end{center}
\end{figure}

On measuring the peaks at $T=7.5$~K (see Fig.~\ref{fig_field_scans_7p5K_Ha}), two field-induced phase transitions have been found.
The transition at 0.2~T is marked by the disappearance of the AFM peaks and also by a sharp increase in the intensity of the FM peaks, which then remain constant up to 2.5~T.
This transition, accompanied by significant hysteresis, is from the AFM to the fully polarised FM state, where a nearly full magnetic moment on both Co$^{2+}$ sites is recovered according to the magnetisation measurements.\cite{Wilson_JPCM_2007}

The measurements of the peak intensity as a function of magnetic field and the $\omega$~scans both show that at the base temperature of 1.6~K, the application of a magnetic field of up to a maximum of 2.5~T along the $a$ axis does not significantly change the intensity of the FM peaks.
No new peaks have been found in applied magnetic fields along the $(0k1)$ and $(0k2)$ lines in reciprocal space.
Neither the position, nor the shape, nor the intensity of the FM peaks such as, for example, $(002)$ and $(022)$ are altered by an applied magnetic field.
This observation suggests that the rapid increase in magnetisation seen at base temperature in small fields (up to 0.1~T)~\cite{Wilson_JPCM_2007} is caused by a rearrangement of magnetic domains.
However, the measured value of the saturated magnetic moment, of about 3.5~$\mu_B$ per Co$^{2+}$ ion, is significantly larger than the average of the zero-field-ordered magnetic moments on spine and cross-tie sites, which are only 2.73~$\mu_B$ and 1.54~$\mu_B$ according to powder neutron-diffraction reports by Chen {\it et al.}~\cite{Chen_PRB_2006} or 3.04~$\mu_B$ and 1.81~$\mu_B$  according to Wilson {\it et al.}~\cite{Wilson_PRB_2007} giving indirect support to the idea of nonzero moments on the vanadium and oxygen ions.\cite{Qureshi_PRB_2009}
The earlier reported gain in the intensity of the FM peaks (and the associated values of the magnetic moment on the cross-tie site) in a powder neutron-diffraction experiment~\cite{Wilson_PRB_2007} is much more gradual in nature, as saturation is reached at about 8~T.

At $T=6$~K, which in zero field corresponds to the $k=\frac{1}{3}$ AFM state, the field dependence of the magnetic peaks demonstrates a steady increase in intensity for the FM type and a gradual intensity decrease for the AFM type.
At this temperature, a relatively weak field, less than 0.2~T, destroys the AFM order and induces the FM state.
Remarkably,  a significant hysteresis is observed in the intensity of the AFM peaks, while the intensity curves for the FM peaks for increasing/decreasing field are practically indistinguishable.

Another transition in a lower field ($\approx 0.072$~T with a significant hysteresis) is clearly marked by a sharp minimum in the intensity of the AFM peaks, as shown in Fig.~\ref{fig_field_scans_7p5K_Ha}, top panel.
The $\omega$~scans performed on the AFM reflection $(0\frac{1}{2}2)$ at 7.5~K in zero field, 0.072~T and 0.15~T have revealed, however, that the observed intensity decrease is mostly due to the movement of the peak to a slightly different position (see inset in Fig.~\ref{fig_field_scans_7p5K_Ha}).
A shift of approximately 0.18$^\circ$ away from the $(0\frac{1}{2}2)$ position suggests that in a field of 0.072~T the magnetic structure is incommensurate with $k \approx 0.475$ instead of the commensurate AFM position with $k=\frac{1}{2}$, although a considerable increase in the peak width implies that there is a large spread in the value of $k$ at the transition point.
\begin{figure}[tb]
\begin{center}
\includegraphics[width=0.83\columnwidth]{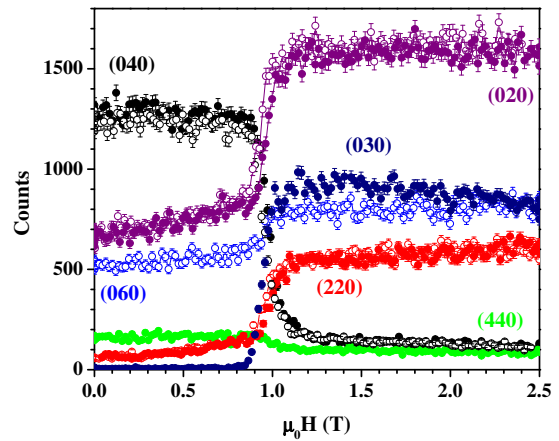}
\caption[]{(Colour online) Magnetic field dependence of the integrated intensity of several peaks in a single-crystal of \CVO\ at $T=1.6$~K  for $H\parallel c$.
                  Solid/open symbols mark increasing/decreasing magnetic field ramps.}
\label{fig_field_scans_baseT_Hc}
\end{center}
\end{figure}

A transition at 0.072~T is also marked by the small change of slope of the intensity of the FM peaks (see Fig.~\ref{fig_field_scans_7p5K_Ha}) and by a change of slope in the magnetisation.\cite{Wilson_JPCM_2007}

\begin{figure}[tb]
\begin{center}
\includegraphics[width=0.83\columnwidth]{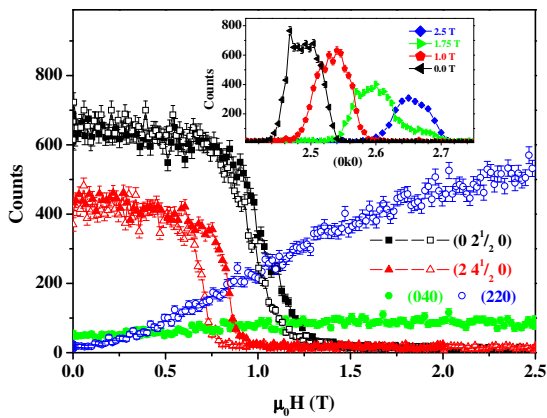}
\caption[]{(Colour online) Magnetic field dependence of the integrated intensity of several AFM and FM reflections in a single-crystal of \CVO\ at $T=7.5$~K  for $H\parallel c$.
                 Solid/open symbols mark increasing/decreasing magnetic field ramps.
                 The inset shows the magnetic field dependence of the neutron-diffraction profiles taken along the $(0k0)$ direction at $T=7.5$~K.}
\label{fig_field_scans_7p5K_Hc} 
\end{center}
\end{figure}

In sharp contrast to the results described above for $H\parallel a$, the application of a magnetic field along the $c$ axis is accompanied by a pronounced change in intensity of the magnetic peaks at all temperatures, including at the base temperature of 1.6~K.
Figs.~\ref{fig_field_scans_baseT_Hc} and~\ref{fig_field_scans_7p5K_Hc} show the field dependence of the intensity of magnetic peaks at $T=1.6$~K and $T=7.5$~K respectively.

At the lower temperature, a pronounced phase transition is seen just below 1~T.
The transition is characterised by an abrupt gain in the intensity of such peaks as $(020)$, $(030)$, $(060)$, $(220)$ and $(420)$ at the expense of such peaks as $(040)$ and $(440)$.
The intensities of the $(020)$ and $(220)$ peaks have a noticeable positive slope in lower fields, which is compatible with the steady rise of the magnetisation.\cite{Wilson_JPCM_2007}
This gradual increase in intensity corresponds to a deviation of the magnetic moments away from their initial orientation parallel to the $a$ axis towards the magnetic field direction along the $c$ axis.
The intensity of the $(030)$ peak is zero in lower fields followed by a sharp increase just below 1~T (see Fig.~\ref{fig_field_scans_baseT_Hc}).
The field variation of the intensity of this peak has a small but distinctly negative slope above 1~T, which indicates that the magnetic structure in this field range has a sizeable \afm ic component and that this component is decreasing steadily in an increasing field, as the magnetic moments become more and more polarised along the field.

The transition into an AFM phase from the initial FM ground state is found to be temperature sensitive, as it shifts from just below 1~T at 1.6~K to nearly 2~T at $T=5.8$~K.
The maximum experimentally available field of 2.5~T has not permitted the observation of a further transition to a fully polarised state, which takes place at 5~T for $H\parallel c$.\cite{Wilson_JPCM_2007}

Unlike the intermediate-temperature zero-field AFM structure characterised by the propagation vector $k=(0\frac{1}{2}0)$, the low-temperature field-induced AFM structure does not involve a doubling of the unit cell.
The observed $(030)$ AFM reflection corresponds to the propagation vector $k=(010)$ (or $(\frac{1}{2}\frac{1}{2}0)$ in a primitive cell) implying that the Co$^{2+}$ atoms located in the adjacent \Kag\ planes carry magnetic moments with significant antiparallel components.
This in turn indicates that the exchange interaction between the planes is strong and \afm ic in nature.
This interaction has so far been neglected in the models proposed to describe the inelastic neutron scattering results.\cite{Ramazanoglu_PRB_2009}
Our results reveal a need for the development of a theoretical model which includes the inter-planar exchange interactions and call for further inelastic neutron scattering experiments.

At $T=7.5$~K the intensity of the integer FM peaks (see Fig.~\ref{fig_field_scans_7p5K_Hc}) is a smooth function of the applied field in the same manner as the overall magnetisation of the sample.\cite{Wilson_JPCM_2007}
The field dependence of the intensity of the AFM half-integer peaks such as $(0\frac{5}{2}0)$ and $(2\frac{9}{2}0)$, however, still demonstrates a pronounced change around 1~T.
This abrupt change in intensity (which is accompanied by a significant hysteresis) corresponds to the change of position of the AFM reflections in reciprocal space rather than to their complete disappearance.
The scans taken along the $(0k0)$ direction in different fields (see inset to Fig.~\ref{fig_field_scans_7p5K_Hc}) show that in higher fields the magnetic structure is incommensurate and the peaks are seen at general positions $(h,\frac{1}{2}+\delta,l)$, where $0<\delta<0.15$, rather than at half integer AFM positions $(h\frac{1}{2}l)$.

These observations encourage an analogy between the effects caused by an applied magnetic field and increasing temperature in \CVO.
In a zero field, a temperature increase induces initially a transition from the ground FM state to the AFM phase and further to the incommensurate phase.
The effects of applying a magnetic field along the $c$ axis are not dissimilar, as it  causes the transition from the FM phase to the AFM phase at lower temperatures and also a transition from the AFM state to the incommensurate state at higher temperatures.

There is an interesting possibility to consider here.
As much as the complex competing interactions are likely to be responsible for the appearance of a large number of magnetic phases in \CVO\ in an applied field, the interactions themselves could be modified significantly by the transitions though the local lattice distortions involving Co centre displacement.\cite{Vergara_PRB_2010}
 
We are grateful to M.~R.~Lees for a critical reading of the manuscript.
We also thank C.~Castelnovo, J.~T.~Chalker and L.C.~Chapon for fruitful discussions.

\end{document}